\documentclass[conference]{IEEEtran}
\IEEEoverridecommandlockouts
\usepackage{cite}
\usepackage{amsmath,amssymb,amsfonts,amsthm}
\usepackage{algorithm}
\usepackage{graphicx}
\usepackage{multicol, blindtext}
\usepackage{textcomp}
\usepackage{svg}
\usepackage{subfig}
\usepackage{amsmath}

\graphicspath{ {figures/} }
\usepackage{color} 
\usepackage{float}
\usepackage{verbatim} 
\usepackage{booktabs}

\usepackage[noend]{algpseudocode}

\def\BibTeX{{\rm B\kern-.05em{\sc i\kern-.025em b}\kern-.08em
T\kern-.1667em\lower.7ex\hbox{E}\kern-.125emX}}

\IEEEoverridecommandlockouts

\begin{document}

\title{\fontsize{23.7}{26}\selectfont
Enabling On-Demand Cyber-Physical Control Applications with UAV Access Points}

\author{
\IEEEauthorblockN{Igor Donevski, Jimmy Jessen Nielsen
\IEEEauthorblockA{Department of Electronic Systems, Aalborg University, Denmark}}
\IEEEauthorblockA{e-mails: igordonevski@es.aau.dk, jjn@es.aau.dk}

\thanks{This work has been submitted to IEEE for possible publication. Copyright may be transferred without notice, after which this version may no longer be accessible.}
} 
\maketitle

\begin{abstract}
Achieving cyber-physical control over a wireless channel requires satisfying both the timeliness of a single packet and preserving the latency reliability across several consecutive packets. To satisfy those requirements as an ubiquitous service requires big infrastructural developments, or flexible on-demand equipment such as UAVs. To avoid the upfront cost in terms of finance and energy, this paper analyzes the capability of UAV access points (UAVAPs) to satisfy the requirements for cyber-physical traffic. To investigate this, we perform a Gilbert-Eliott burst-error analysis that is analytically derived as a combination of two separate latency measurement campaigns and provide an upper-bound analysis of the UAVAP system. The analysis is centered around a UAVAP that uses its LTE connection to reach the backhaul, while providing service to ground nodes (GNs) with a Wi-Fi access point (AP). Thus, we combine both measurement campaigns to analyze the plausibility of the described setup in casual, crowded or mixed network settings.

\end{abstract}

\begin{IEEEkeywords}
Unmanned Aerial Vehicle, Drones, Access Point, Low Latency, Survival Time, Burst Errors
\end{IEEEkeywords}

\section{Introduction}
Connected robotics is a prime driver in the development of the sixth generation (6G) of mobile networks \cite{6Gvision}. This comes as an extension to one of the main service expected from the fifth generation of mobile networks (5G) named Ultra-Reliable and Low-Latency Communication (URLLC) \cite{popovski2019wireless}. Offering URLLC is difficult due to the combined requirement of achieving latency in the order of few milliseconds with reliability of around $1-10^{-5}$. However, the URLLC use case is concerned with latency-reliability of each separate packet which does not capture the operation of multi-packet operations such as cyber-physical control applications. These types of applications transmit regular updates that are not gravely affected by a single untimely packet (a packet that violates the latency requirements). Therefore, cyber-control applications are dependent on the time on which they can operate nominally without having received an eligible packet, referred to as  \emph{survival time}~\cite{3gppvertical}. Therefore, in this paper we focus on analyzing experimentally derived insights on the potential application of Unmanned Aerial Vehicles (UAVs), colloquially referred to as drones. The analysis is targeted at improving communication system reliability through reducing the likelihood of error bursts such as consecutive untimely packets for ground nodes (GNs).

\subsection{Goals \& Motivation}
\begin{figure}[t!]
    \centering
\includegraphics[width=1\columnwidth]{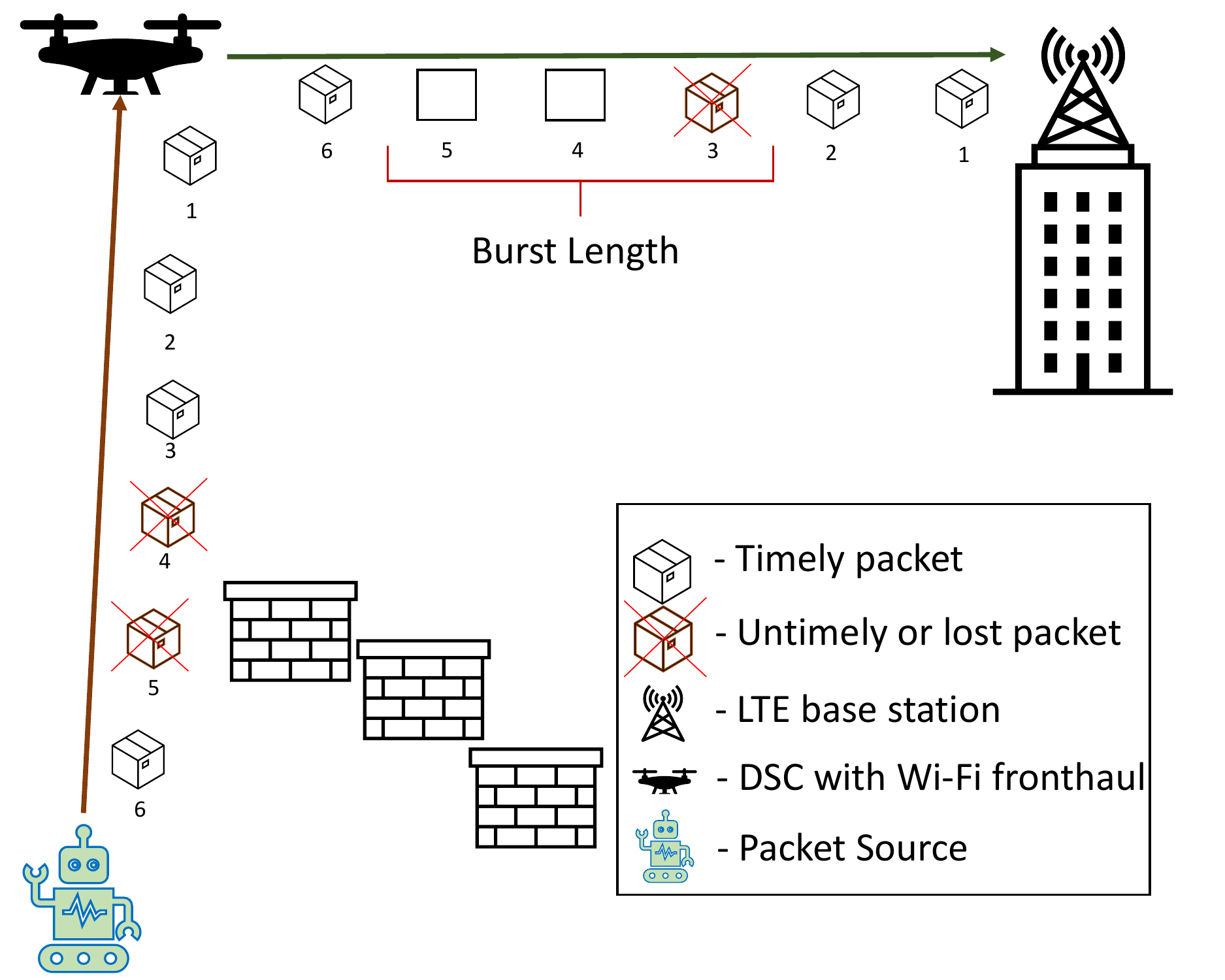}
    \caption{An illustration of the scenario investigating the probability of lengthy error bursts in the use if a UAVAP.
    }
    \label{fig:illustration}
\end{figure}
UAV Access Points (UAVAPs) are a perfect adaptable solution for addressing gaps in wireless communication coverage, mainly as a benefit to their mobility as a modular mobile service equipment \cite{9868834}. This aims to save large quantities of money and energy through avoiding constant operation of Base Stations for this particular service \cite{9857897,mainsurvey}. Moreover, the adaptable horizontal position and altitude, improve the chances of having a Line of Sight both towards base station (BS) towers mounted on top of built up terrain, as well as GNs \cite{tutorial,igoreucnc}. Thus, UAVAPs are an intuitive solution for providing reliable service to GNs with low latencies on both the GN access link (GNAL) and BS access link (BSAL). Therefore, this work intends to address an interest in modeling the error burst probability for a full stack of BSAL and GNAL, where both connections are offered by widely available technologies such as LTE and Wi-Fi respectively.

Despite the high likelihood for LoS on both BSAL and GNAL, both interfaces might provide significant delays when the network conditions get crowded, and channel competition is high. Thus, the goal is to frame the issue of an UAVAP enabling ground based closed-loop end-to-end control applications in different network conditions, by focusing on two key questions. Does UAVAP provided wireless access offer improved latency-reliability in relaxed and crowded GNAL and/or BSAL? What are the shortcomings of the UAVAP wireless access and how can those be addressed? 

The paper organization is done in the following manner. Section~\ref{sec:literature} discusses significant works in the fields of cyber-physical control, burst error analysis, and experimental UAVAP systems. Section~\ref{sec:berr} introduces the model for burst errors in UAVAP systems. Section~\ref{sec:preana} showcases statistical properties of the considered measurement campaigns. Section~\ref{sec:results} showcases the burst error performance of a UAVAP system. Finally, Section~\ref{sec:conclusion} summarizes the results of the analysis, addresses concerns, and discusses possible future directions.

\section{Related Work}
\label{sec:literature}
The experimental and theoretical communities are generally concerned with several complementary problems of UAVAP systems, particularly for addressing reliability concerns. The work of \cite{gangula2018flying} does a throughput investigation for an autonomously repositioning UAV in a highly controlled scenario. We previously investigated how the UAV position can be exploited when having a fixed coverage area \cite{igoreucnc}, and upgraded the analysis for wireless network slicing \cite{igorwcnc}.  Nonetheless, analyses that focus on the positioning of a UAVAP are generally concerned with optimizing the wireless conditions of the GNAL, and rely on some strong assumptions for the BSAL \cite{mainsurvey}. 

There are experimental efforts to address some more common problems that stem from the design of cellular systems for UAVAPs. As covered in the overview in \cite{lin2019mobile}, the two central issues that appear in such systems are the higher inter-cell interference from distant BSs and the issue of antenna propagation design that is tilted toward the ground users. Although these issues add to lessen the reliability of the system, they introduce frequent handover between the close and distant BSs, and are covered in several different measurement campaigns, and are well elaborated in the following works \cite{fakhreddine2019handover,nguyen2018ensure,van2016lte}. However, there was a lack of experimentally derived latency for UAVAPs, which was thus covered in our previous work \cite{igorlakeside}.

The survival time of cyber-physical control applications varies depending on the application and can be generally inferred from the report in \cite{3gppvertical}. In accord, the work of \cite{DeSantAna2020} observes that  the safety of autonomously guided road vehicle systems is directly reliant on the number of consecutive errors. Our previous work in \cite{donevski2021performance} analyzed the efficiency of multi-connectivity for servicing cyber-physical control traffic. As such, this paper addresses a gap in the literature for burst error analysis for UAVAPs by building a model for analyzing temporal correlations in each interface, for two different sets of network conditions. 

\section{Burst Errors in UAVAP systems}
\label{sec:berr}


In an end-to-end link between a GN and a central controller, data is transferred periodically with an occurrence of $T_s$. 
Packet timeliness must be satisfied (packet latency must not exceed some value $\theta\leq T_s$ ) with the goal of keeping all actions taken by the robot fresh and not outdated. Therefore, just like in the URLLC use case we look at the latency-reliability function (a CDF of the system latency) \cite{Nielsen2018}. Moreover, since our communication occurs on two different links we identify the total UAVAP latency limit as the sum of both $\sum_i \theta_i < \theta$, where $i \in \{ \text{Wi-Fi}, \text{LTE} \}$ for the GNAL and BSAL respectively. 

The total latency budget $\theta$ needs to be split for each interface. For a latency of $l_i$ on interface $i$, and its latency deadline $\theta_i$ we define the latency-reliability function as:
\begin{equation}
    F_i(\theta_i) = \Pr(l_i\leq \theta_i).
\end{equation}
Thus, we define a packet error in contrary to traditional cases where $\theta\rightarrow \infty$ but as the case where a single of both interfaces is untimely as in:
\begin{equation}
    P_e^{(i)}=1- F_i(\theta_i).
\end{equation}
Treating each link individually like this, means that we judge the UAVAP system by its upper bound of performance where neither of the links has violated its timeliness, regardless of the final collective $\theta$. This allows for bad states of each interface to be identified quickly and treated individually in the two hops of communications that need to be nominal for a packet to be timely. 

\begin{figure}[t!]
    \centering
\includegraphics[width=0.6
\columnwidth]{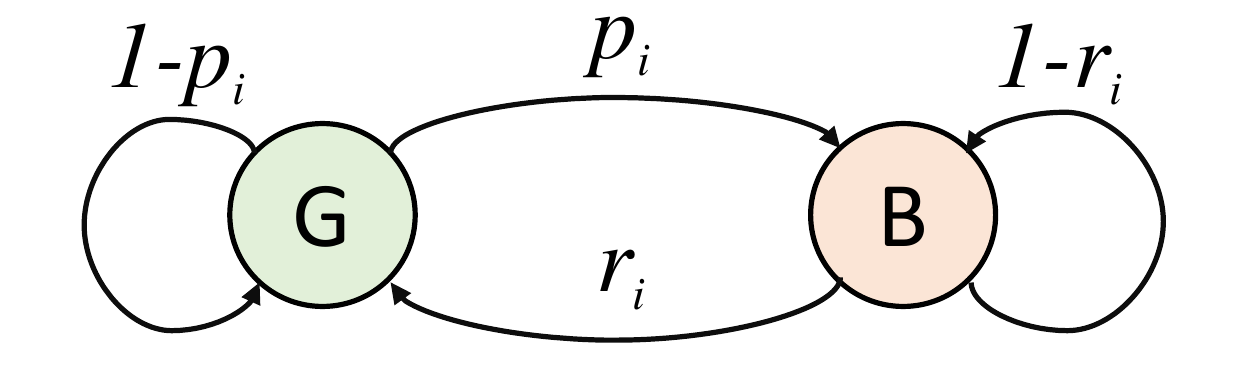}
    \caption{The GE system for modeling burst errors.
    }
    \label{fig:GEtrad}
\end{figure}

Therefore, each untimely transmission on each interface of the UAVAP can be considered as a transition to a failed state for that link of the hop. In this way, the complexity of the collective wireless conditions such as congestion, small/fast fading, noise and interference are collected and simplified to two states ( a \emph{good} state G and a \emph{bad} state B, as illustrated in Fig.~\ref{fig:GEtrad}) of a discrete time Markov chain. This model is the Gilbert-Elliott (GE) burst error model that has two parameters $p_i$ and $r_i$ that are the probabilities to transition from the good to the bad state and from the bad state to the good one, respectively \cite{donevski2021performance}. Thus to model the GE system accurately, we require statistically relevant data that define the state transitions in \eqref{eq:GEmatrix} accurately.
\begin{equation}
\mathbf{P}_i = 
 	\begin{bmatrix}
    1-p_i & p_i \\
    r_i& 1-r_i
  \end{bmatrix},
  \label{eq:GEmatrix}
  \end{equation}
with steady states:
\begin{align}
\label{eq:PiB}
	\pi_{i,\text{G}} = \frac{r_i}{p_i+r_i}, \\
	\pi_{i,\text{B}} = \frac{p_i}{r_i+p_i}, 
\end{align}
and total probability of the system must satisfy $\pi_{i,\text{G}} + \pi_{i,\text{B}} = 1$.

\begin{figure}[t!]
    \centering
\includegraphics[width=0.6\columnwidth]{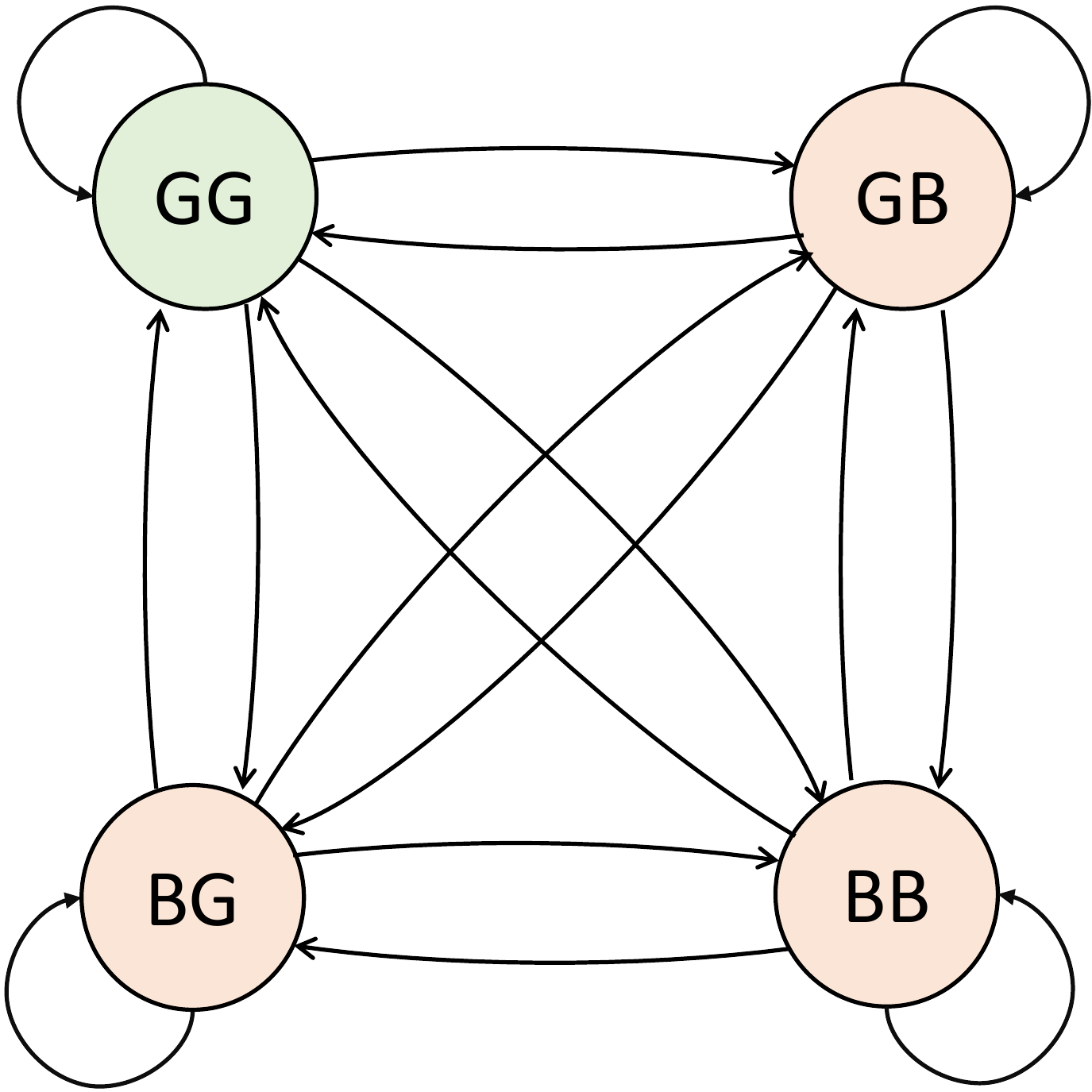}
    \caption{The adapted GE system for modeling burst errors across a UAVAP with independent GNAL and BSAL.
    }
    \label{fig:4states}
\end{figure}

Therefore, given the GE model for each interface, we can convert the total end-to-end link as a four state GE model as shown in Fig.~\ref{fig:4states}. Here, each interface moving over to a bad state results into a bad overall state and compromises the total link.

\section{Measurement Campaigns}
\label{sec:preana}
In this section we showcase and compare two different testing campaigns on Wi-Fi and LTE latencies. The two sets of probabilities that are contained in the subsections that follow are taken as, representative of direct crowded links in an NLoS scenario (Subsection \ref{sec:crowded}), relaxed radio conditions in a purely LoS scenario (Subsection \ref{sec:relaxed}), and relaxed radio conditions for a direct GN to BS link (Subsection \ref{sec:drelaxed}).
\subsection{Crowded Scenario}
\label{sec:crowded}
\begin{figure}[htb]
\centering
\includegraphics[width=1\columnwidth]{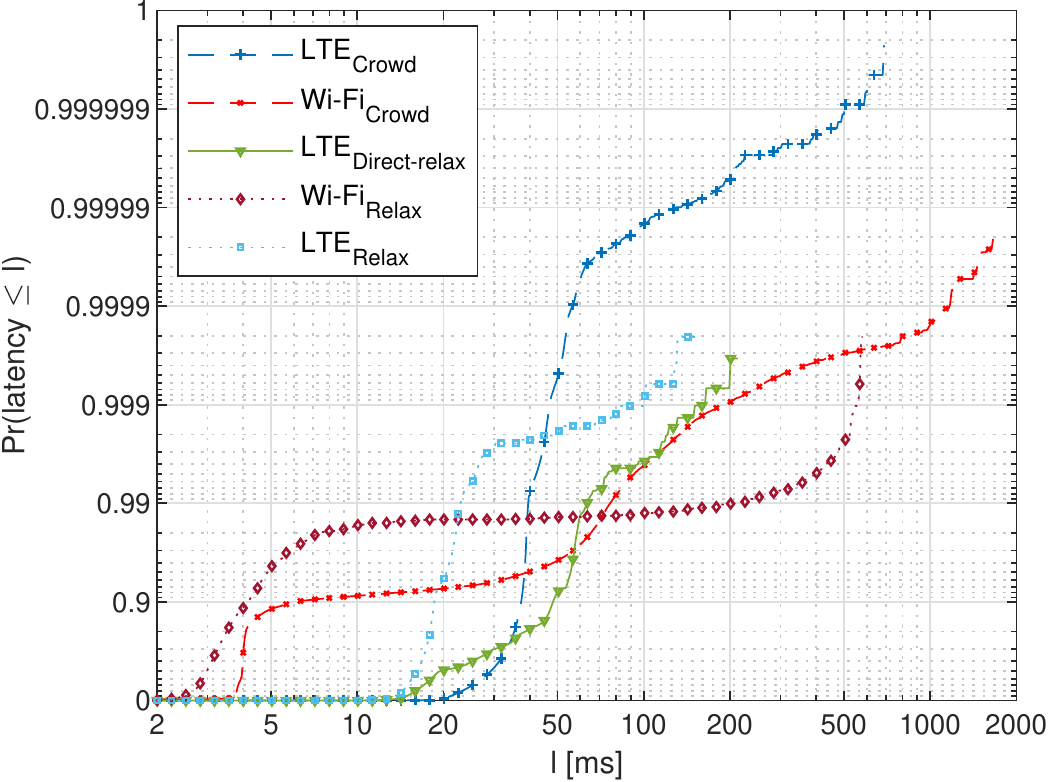}
\caption{Empirical latency CDFs for \textbf{Relaxed}, \textbf{Crowded}, and \textbf{Direct-relax}.
}
\label{fig:latrel}
\end{figure}
The initial trace was taken in a challenging environment that was crowded with competing devices, and is thus referred as the \emph{Crowd} scenario. This was a trace of latency measurements of sending small (128~bytes) UDP packets every 100~ms between a server and a node that connects to an LTE BS, or Wi-Fi AP, at the Aalborg University campus \cite{donevski2021performance}. These measurements were taken in a fully NLOS (indoor) environment with physically-static devices that had a varying channel competition during the day. Nonetheless, the measurements were taken in good-SNR radio conditions, which enhances the differences between LTE and Wi-Fi protocol operation in the presence of competing traffic. Most importantly, the medium access protocol (MAC) varies on whether it needs to operate in some licensed spectrum (for LTE), or it needs to operate in a highly contentious unlicensed spectrum (for Wi-Fi).
In Fig.~\ref{fig:latrel} we can see that Wi-Fi can provide 5~ms delays for 90\% of its packets. However, if an application allows for more relaxed timeliness constraint, 99\% of packets arrive before the 80~ms deadline. On the other hand, it is very hard to achieve very low latency for LTE, even for a very small portion of packets. Therefore there is little difference in the LTE latency deadlines, 36~ms and 40~ms, for 90\% and 99\% of packets, respectively.

\subsection{Relaxed Scenario}
\label{sec:relaxed}

The second trace was taken in a more controlled scenario where the contention for the wireless resources was relaxed and is thus referred to as the \emph{Relax} scenario. The measurements are taken outdoors, with full LOS links, in the 5G Playground Carinthia testbed in the outskirts of Klagenfurt, Austria. This was a full UAVAP implementation where the samples were taken while flying along a pre-planned path with a constant speed of 1~m/s. Each 100~ms a small packet (20 byte payload) was generated and transmitted from a ground user, sent through the UAVAP to the LTE BS. The testbed was set in a way that is easy to extract and separate the LTE and Wi-Fi behavior\cite{igorlakeside}.

From the statistical performance of Wi-Fi and LTE In Fig.~\ref{fig:latrel} we can notice that both interfaces retain the same statistical behaviour as in the Crowd scenario but with different timing performances. In general, the measured latencies were slightly lower due to the smaller physical distance of the server from the LTE BS. The Wi-Fi can provide 5~ms timeliness to 95\% of the packets, while the 99\% packet pass-through rate is reached when the timeliness constraint is set to be 200~ms. The LTE connection here has a significantly wider latency gap for 90\% and 99\% of packets, 18~ms and 23~ms, respectively.

\subsection{Direct-relaxed Scenario}
\label{sec:drelaxed}
Within the same \emph{Relax} testing scenario in Klagenfurt, a separate measurement trace was taken only for the GN that directly connects to the BS. This was to provide realistic and proportional measurement set to compare against the UAVAP setup. In this case, the GN had mixed LOS and NLOS LTE link with the BS along the same outdoor path and speed taken in the previous trace. This is common outdoor environment where reliable implementation of cyber-physical applications is needed. Finally, we note that this measurement is strongly correlated with the LTE$_\text{relax}$. This correlation can be exploited to generate a synthetic GE model for a LTE$_\text{direct-crowded}$

\begin{table}[t]
\begin{center}
\caption{$p$ and $r$ transition probabilities used in evaluation.}
\label{tab:pnr}
\begin{tabular}{lccccc}\toprule
      & \multicolumn{2}{c}{Crowd} &  & \multicolumn{2}{c}{Relax} \\ \cmidrule(lr){2-3} \cmidrule(lr){5-6}
      & $p$         & $r$         &  & $p$         & $r$         \\ \cmidrule(lr){1-6}
Wi-Fi & 0.0515        & 0.9468     &  & 0.0001  & 0.0845        \\
LTE   & 0.0178        & 0.2577     &  & 0.0127  & 0.8356 \\      \bottomrule
\end{tabular}
\end{center}
\end{table}

\section{Numerical Analysis \& Results}
\label{sec:results}
Combining latency measurements is a difficult feat due to the vast variability of each testing setup, unpredictable environment, unforeseen biases, and traffic proportions that result in incompatible latency distributions. However, the GE model can abstract the complexity of the aforementioned challenges by setting an adequate $\theta$ that well represents the proportions of each respective testing setup. With this, we want to achieve a thorough analysis of the burst error performance in any combination of crowding on each links.

Since both measurement campaigns concern vastly different properties and BSAL networks, the timing constraints had to be scaled, so that the results can be comparable. We have done this by considering the equi-performance points for both Wi-Fi and LTE. In detail, these are the delay values for which both interfaces, LTE and Wi-Fi provide the same reliability within their respective measurement setup, or simply, the point where the CDF lines cross. For the crowded scenario the delay separating a bad versus a good state of the channel is $\theta_\text{crowd} = 38.25\,\text{ms}$. For the relaxed scenario we set the cutoff delay per hop to $\theta_\text{relax} = 22.5\,\text{ms}$. Since the interfaces produce similar behaviour in both measurement campaigns, the crossing point is the only relevant point that is correlated between both campaigns. Setting the delay values like this gives a balanced weight to both interfaces given the two different testing environments, and allows us to analyze bursty losses across multiple packets. This isolates the GE models for each interface and measurement campaign separately, given the different distances to the server. In the case of the two different measurement campaigns, the baseline latency difference is 0.7 times larger for the \emph{Crowd} setup. Therefore, the resulting $p$ and $r$ values are presented in Tab. \ref{tab:pnr}. 

To compare the performance of the UAVAP we use the LTE-direct-relaxed connection for a GN outdoors and on the ground. Since the Relax and the LTE-direct measurements are taken in the exact same testing conditions, we used the following reasoning for the latency deadline $\theta_\text{direct-relaxed} = 2 \cdot \theta_\text{relax} = 45~\text{ms}$. This is a best case scenario for the $\theta_\text{direct-relaxed}$ as it is equivalent to the upper performance bound where not one, but both interfaces are in a bad state in the UAVAP setup. The resulting GE model values for this measurement scenario are $p_\text{direct-relaxed} = 0.1457$ and $r_\text{direct-relaxed} = 0.7857$. Following this, we used the LTE$_\text{crowd}$ measurements to synthesize the $p$ and $r$ values for a potential LTE$_\text{Direct-crowd}$ by following the proportional correlation between LTE$_\text{relax}$ and LTE$_\text{Direct-relax}$. This resulted in the $p_\text{direct-crowd} = 0.2042$ and $r_\text{direct-crowd} = 0.2423$ values for the fictional LTE$_\text{Direct-crowd}$ dataset.
    
\begin{figure}[t]
\centering
\includegraphics[width=1\columnwidth]{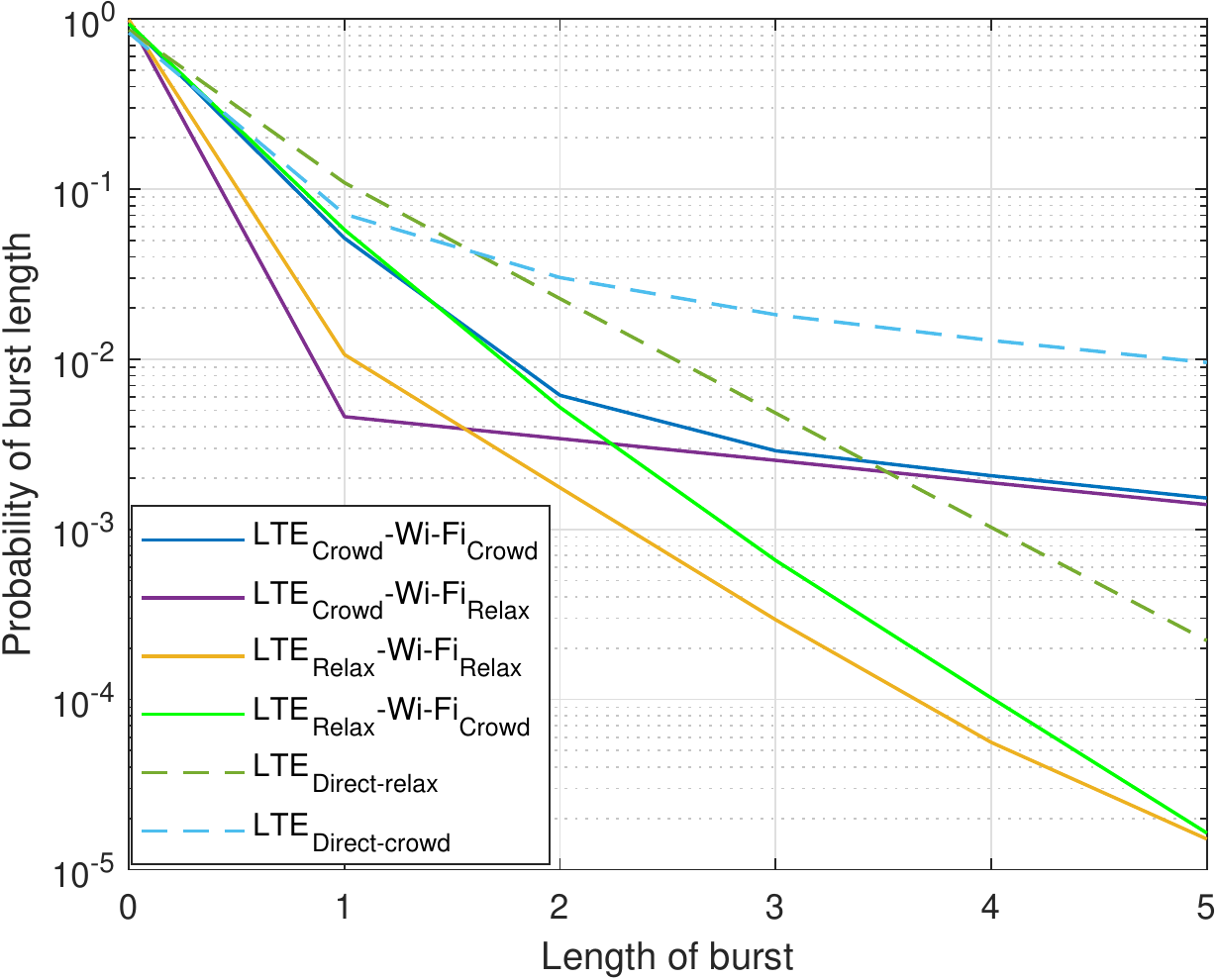}
\caption{Burst error length analysis for all combinations of possible environments \emph{Crowd} and \emph{Relax} for each interface.
}
\label{fig:latrel1}
\end{figure}

An analysis between both sets of measurements are that the LTE $p$ probabilities are almost equal, with $p_\text{LTE relax}$ having a slightly smaller probability to move over to the bad state. Moreover, there is a larger discrepancy between the LTE $r$ values in both environments, where $r_\text{LTE relax}$ has a much larger probability to return to the good state once it has performed poorly. This makes a big difference in the steady state probability of being in the bad channel state of 0.0150 for relaxed, versus 0.0646 for the crowded.
There are clear differences between the performance of the LTE and Wi-Fi interfaces, however, as per the previous section the issue of using UAVAP is that the negative effects of both interfaces are stacked when relaying. 

There are more distinct differences between both sets of Wi-Fi measurements. While the probability for the Wi-Fi to go to the bad state are generally small $p_\text{Wi-Fi relax}$ is orders of magnitude lower than in the crowded scenario. However, the probability of the Wi-Fi connection to recover in the crowded scenario is much higher than in the relaxed one. This is the case because outages in the relaxed scenario are due to rare but persistent channel outages. Therefore, the outage probability of the relaxed Wi-Fi has a value of 0.0015, versus the outage probability of crowded Wi-Fi of 0.0516.

In Fig.~\ref{fig:latrel1} we show results of a Monte-Carlo simulation, of $10^9$ samples, for all possible combinations of the measurement campaigns. It is common that cyber-physical control applications for the GN to allow for error bursts of 3 consecutive packet errors \cite{donevski2021performance,3gppvertical} (equivalent to 300~ms). With such an implementation in mind we can notice that all UAVAP implementations outperform a direct-relaxed LTE link. Moreover, the usefulness of the UAVAP is not very strong when the LTE connection is in the crowded mode. However, this would mean that the LTE link for the direct implementation.

\begin{figure}[t]
\centering
\includegraphics[width=1\columnwidth]{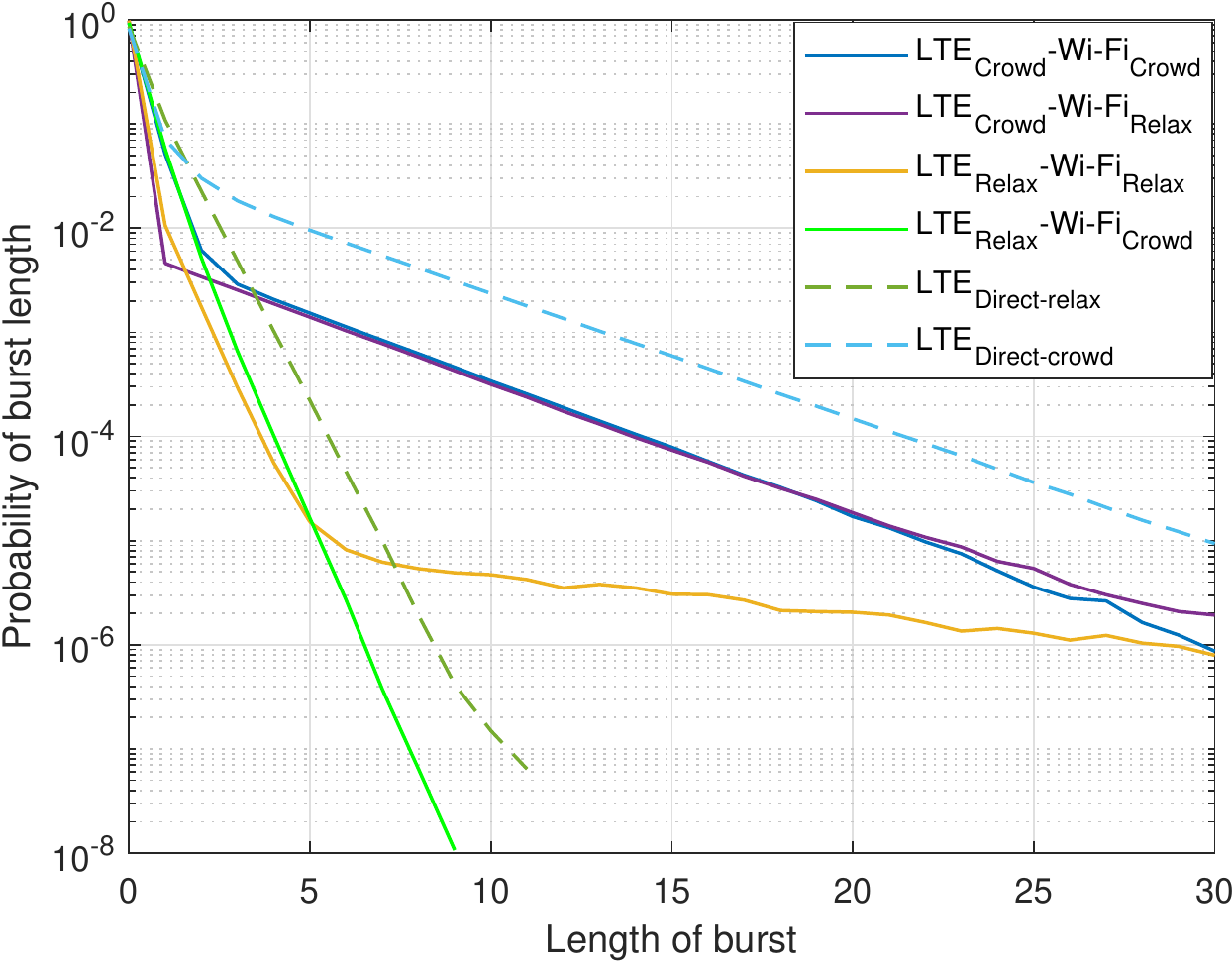}
\caption{Burst error length analysis for all combinations of possible environments \emph{Crowd} and \emph{Relax} for each interface.
}
\label{fig:latrel2}
\end{figure}

In Fig.~\ref{fig:latrel2} we expand the investigation to longer burst length tolerances, up to 30 samples (equivalent to 3~s). We can see that the low probability of Wi-Fi to exit a bad state has a bad influence on the system. This is particularly visible in the LTE$_\text{relax}$-Wi-Fi$_\text{Relax}$ scenario that rarely went into a bad state, but remains there for a very long time. This "flattens" the  burst length performance when poor SNR conditions, or other systematic failures that affect Wi-Fi occur. 

\section{Conclusions}
\label{sec:conclusion}

In this work we analyzed the burst packet error probability of UAVAPs for providing on-demand communications services for the purpose of cyber-physical applications. We did this by using the GE 2-state model for simulating temporal correlation in wireless channels from two separate sets of latency measurements of LTE and Wi-Fi. The two different sets of latency measurements were subjected to different latency deadline values as a way to demonstrate the timeliness sensitivity of packets. The two sets, one taken in a busy spectrum and another in a relaxed spectrum are used to provide a more generalized picture from the initial measurements. The results show that using a UAVAP in order to avoid burst packet errors was always superior to transmitting to a distant LTE BS directly. The advantages can vary depending on the crowding of the radio resources. 

We need to note that even though this analysis contains varied measurement data, further investigation is needed to verify the usefulness of UAVAPs for ubiquitous deployment. In particular, the LTE connection to the main grid can drastically vary on the deployed environment. Finally, this work inspires future studies where the GN exploits both interfaces, such as in the case of multi-connectivity, to exploit the advantages of both the direct and the UAVAP links.

\section*{Acknowledgment}
The work was supported by the European Union's research and innovation programme under the Marie Sklodowska-Curie grant agreement No. 812991 ''PAINLESS'' within the Horizon 2020 Program.

\bibliographystyle{IEEEtran}
\bibliography{./bibliography.bib}

\end{document}